\documentclass{midl} 

\jmlrvolume{-- Under Review}
\jmlryear{2025}
\jmlrworkshop{Full Paper -- MIDL 2025 submission}
\editors{Under Review for MIDL 2025}

\title[Analytical variability with style transfer]{Mitigating analytical variability in fMRI with style transfer}

\midlauthor{\Name{Elodie Germani\nametag{$^{1}$}} \Email{elodie.germani@ukbonn.de}\\
\Name{Camille Maumet\midljointauthortext{Joint senior authorship}\nametag{$^{1}$}} \Email{camille.maumet@inria.fr}\\
\Name{Elisa Fromont\midlotherjointauthor\nametag{$^{2}$}} \Email{elisa.fromont@irisa.fr} \\
\addr $^{1}$ Univ Rennes, Inria, CNRS, Inserm, Rennes, France \\
\addr $^{2}$ Univ Rennes, IUF, Inria, CNRS, Rennes, France
}

\begin{document}

\maketitle

\begin{abstract}
We propose a novel approach to facilitate the re-use of neuroimaging results by converting statistic maps across different functional MRI pipelines. We make the assumption that pipelines used to compute fMRI statistic maps can be considered as a style component and we propose to use different generative models, among which, Generative Adversarial Networks (GAN) and Diffusion Models (DM) to harmonize statistic maps across different pipelines. We explore the performance of multiple GAN and DM frameworks for unsupervised multi-domain style transfer. We developed an auxiliary classifier that distinguishes statistic maps from different pipelines, allowing us to validate pipeline transfer, but also to extend traditional sampling techniques used in DM to improve the transition performance. Our experiments demonstrate that our proposed methods are successful: pipelines can indeed be transferred as a style component, providing an important source of data augmentation for future studies. \end{abstract}

\begin{keywords}
style transfer, generative models, analytical variability, fMRI, data re-use
\end{keywords}

\section*{Introduction}\label{intro}

Over the last decades, the question of understanding brain functions has taken an important place in many research fields. With the development of brain imaging techniques such as task-based functional MRI (task-fMRI), researchers can now explore the brain activity of individuals while they perform predefined tasks, and get a better understanding of the neural correlates of different cognitive processes. 

However, the ``reproducibility crisis'' raised concerns about the reliability of published findings, including in neuroimaging~\citep{button_power_2013,poldrack_scanning_2017,botvinik-nezer_reproducibility_2023}, and prompted the adoption of new research practices. Efforts have been made to increase sample sizes by acquiring data from a larger number of participants for a few numbers of cognitive tasks (\textit{e.g.} UK Biobank~\citep{sudlow_uk_2015}) or for a small number of participants on a larger number of cognitive tasks (e.g. Individual Brain Charting~\citep{pinho_individual_2018}). However, the number of research questions that can be explored is always limited by the characteristics of each dataset. 

With the increased adoption of data sharing~\citep{poline_data_2012}, more and more neuroimaging data are made available on dedicated platforms (\textit{e.g.} OpenNeuro~\citep{markiewicz_openneuro_2021} or NeuroVault~\citep{gorgolewski_neurovaultorg_2015}). Re-using shared data in new studies would allow researchers to explore new research questions, with larger and more diverse datasets, while bypassing the difficulties associated with acquiring new data. Derived data could also be combined instead of raw data through meta- and mega-analyses~\citep{costafreda_pooling_2009}. This process also reduces privacy constraints and avoids costly re-computations. 

In task-fMRI, due to the high flexibility of analytical pipelines~\cite{carp_plurality_2012}, derived data shared on public databases often come from different pipelines. However, different pipelines can lead to different results~\cite{botvinik-nezer_variability_2020}, and combining results from different pipelines in mega-analyses can lead to a higher risk of false positive findings~\cite{rolland_towards_2022}. To benefit from this large amount of derived data available, it is thus necessary to find a way to mitigate the effect of pipeline differences on derived data.

To mitigate the effect of different sources of variability, researchers usually perform data harmonization with techniques such as style transfer~\citep{gatys_image_2016} that allow learning mappings between different domains. This technique makes use of generative models such as Generative Adversarial Networks (GANs)~\cite{goodfellow_generative_2014} and Diffusion Models (DMs)~\cite{ho_denoising_2020}. Supervised frameworks~\cite{isola_image--image_2018,saharia_palette_2022} can be trained to learn a mapping between data pairs from different domains. In contrast, unsupervised frameworks~\cite{zhu_unpaired_2020,sasaki_unit-ddpm_2021,liu_unsupervised_2018} do not necessitate pairs of data in different domains for their training as they use constraints like cycle consistency~\cite{zhu_unpaired_2020} or shared latent space assumption~\cite{liu_unsupervised_2018,sasaki_unit-ddpm_2021}. These provide a good opportunity to benefit from large unlabeled databases and learn complex mapping without ground truth. By conditioning on domain-specific features, unsupervised frameworks also extend to multi-domain~\cite{choi_stargan_2018,choi_ilvr_2021,ho_classifier-free_2022} to learn transfer between multiple domains in a single model. 

In this work, we explore the ability of style transfer frameworks to convert task-fMRI statistic maps between pipelines. Our goal is to propose a solution to mitigate the effect of pipeline differences for fMRI mega-analyses in order to benefit from the large amount of derived data shared on public databases. To be useful in real practice, the proposed method should rely on unpaired data and perform multi-domain transitions. To our knowledge, this application of style transfer to data conversion between different analysis pipelines is new, and off-the-shelf methods do not directly apply as these were not designed on the same type of data. 

To tackle these challenges, we made the following contributions:  
\begin{itemize}
    \item We are the first to make the assumption that pipelines can be considered as a style property of statistic maps which can be transferred between maps. 
    \item We re-implement three state-of-the-art style transfer frameworks based on GAN, namely Pix2Pix~\citep{isola_image--image_2018}, CycleGAN~\citep{zhu_unpaired_2020} and StarGAN~\citep{choi_stargan_2018}, and adapt them to our 3-dimensional statistic maps. 
    \item We re-implement a state-of-the-art conditional DM~\citep{ho_classifier-free_2022} and adapt it for style transfer by conditioning the sampling on the source image. 
    \item We explore different types of conditioning for DM frameworks: in particular using the latent space of an auxiliary classifier trained to distinguish statistic maps between pipelines, a task previously unexplored.
\end{itemize}

\section*{Materials and Methods}\label{styletransfer:matmet}

\subsection*{Dataset}\label{styletransfer:matmet:dataset}
We use group-level statistic maps from the \textit{HCP multi-pipeline dataset}. More details can be found in~\cite{germani_hcp-multi-pipeline_2024}. Briefly, this dataset is composed of subject-level (1,080 participants) and group-level (1,000 groups) statistic and contrast maps derived from raw data of the \textit{Human Connectome Project Young Adult} S1200 release~\citep{van_essen_wu-minn_2013}. In this dataset, raw fMRI data for the motor task were analyzed with 24 different pipelines for the 5 contrasts: \textit{right-hand}, \textit{right-foot}, \textit{left-hand}, \textit{left-foot} and \textit{tongue}. The pipelines used in this dataset vary in terms of software package, smoothing kernel Full-Width at Half-Maximum (FWHM), number of motion regressors, and derivatives of the Haemodynamic Response Function (HRF) included in the first-level analysis.

We explore in particular the statistic maps obtained with four different pipelines that differ in terms of software packages (SPM~\citep{penny_statistical_2011} or FSL~\citep{jenkinson_fsl_2012}) and the presence or absence of the derivatives of the HRF for the first-level analysis. We use all the available group-level statistic maps (\(N=1,000\)) for each pipeline for the contrast \textit{right-hand}. In the following, these pipelines are labeled with ``$<$software$>$-$<$derivatives$>$'', for instance ``fsl-1'' means the use of FSL software package with HRF derivatives.

The selected group-level statistic maps are resampled to a size of 48~x~56~x~48 and masked using the intersection mask of all groups. The voxel values are normalized between -1 and 1 for each statistic map using a min-max operation. The 1,000 groups are split into train and test with an 80/20 ratio and all models are trained and evaluated on the same sets. Further investigation about possible data leakage across groups is provided in Supplementary Figure 1~\citep{germani_2024_13748563}.

\subsection*{GAN frameworks}\label{matmet:gan-frameworks}

First, we assess the potential of GAN frameworks to convert statistic maps between pipelines. In particular, we evaluate the performance of Pix2Pix~\citep{isola_image--image_2018}, CycleGAN~\citep{zhu_unpaired_2020} and StarGAN~\citep{choi_stargan_2018}. A detailed description of each framework is available in the corresponding papers. We provide a quick description of the main properties of these models in Table~\ref{tab:styletransfer:matmet:gan-frameworks}. 

\begin{table}[htbp!]
    \centering
    \begin{tabular}{p{5.1cm}|p{2.3cm}|p{2.5cm}|p{2.5cm}}
        \hline
        Framework & Learning & Transition & Loss\\
        \hline \
        Pix2Pix~\citep{isola_image--image_2018} & Supervised & One-to-one & Adversarial \newline Reconstruction \\
        CycleGAN~\citep{zhu_unpaired_2020} & Unsupervised & One-to-one & Adversarial \newline Cyclic \\
        StarGAN~\citep{choi_stargan_2018} & Unsupervised & Multi-domain & Adversarial \newline Cyclic \newline Classification \\ 
        \hline
    \end{tabular}
    \caption{Description of GAN frameworks}
    \label{tab:styletransfer:matmet:gan-frameworks}
\end{table}

\paragraph{Architecture and training}
We use the default architecture of these models, as described in their respective papers, and we only modify the 2-dimensional convolutions and batch normalization layers to cope with our 3-dimensional statistic maps. These were implemented using PyTorch~\citep{pytorch_2019} and each framework was trained for 200 epochs on 1 GPU NVIDIA Tesla V100. 

\subsection*{DM frameworks}

Due to the promising performance of DM on natural images and medical imaging~\citep{dhariwal_diffusion_2021}, we also assess the potential of DM frameworks. However, only a few DM frameworks have been developed for style transfer, and to our knowledge, they rely on paired data~\citep{saharia_palette_2022} or learn only one-to-one transitions~\citep{pan_cycle-guided_2023}. Thus, to perform multi-domain transitions, we adapt an existing conditional DM to style transfer tasks. We use the framework from~\cite{ho_classifier-free_2022}, which generates images conditioned using a one-hot encoding of the class (\textit{i.e.} class vector). We extend this model to conditioning based on the latent space of an auxiliary classifier, inspired from~\cite{preechakul_diffusion_2022}. Both are unsupervised frameworks, learning multi-domain transitions. A more detailed description of the original framework is available in~\cite{ho_classifier-free_2022}. 

\paragraph{Source content preservation} To adapt our framework to style transfer, we first needed to find a solution to generate images that maintain the intrinsic properties of the source image. In~\cite{saharia_palette_2022}, authors concatenate the source image with random Gaussian noise to initialize the diffusion. Here, we propose to fix the initial state of the DM by directly using the forward diffusion process to generate a noisy version of the source image \(X_t\). Then, the noisy source image is iteratively denoised using the predicted noise conditioned on the target domain and the reverse diffusion process. 

\paragraph{Classifier conditioning} We also extend the model from~\cite{ho_classifier-free_2022} with conditioning on the latent space of an auxiliary classifier. In~\cite{ho_classifier-free_2022}, the diffusion is conditioned using a one-hot encoding of the domain, limiting the diversity of generated samples. In~\cite{preechakul_diffusion_2022}, a semantic encoder is used to guide sampling. We extend this idea by conditioning the model on a latent feature vector extracted from a pre-trained CNN. This CNN was pre-trained to distinguish pipelines from the statistic maps. The features are extracted just before the classification layer, to get a good representation to distinguish images across pipelines.

\paragraph{Multi-target images} \cite{choi_ilvr_2021} showed that conditioning on multiple images generates images that share more coarse or fine features with the target ones depending on the number of selected images. Here, we aim to better represent the heterogeneity of the target domain. In practice, the whole set of images available in the target domain could be used. This is impractical for large datasets and might lead the model to focus on specific patterns of the target domain if these are over-represented in the dataset. 

\paragraph{Architecture and training}
The neural network used to predict the noise follows a simple U-Net architecture~\citep{ronneberger_u-net_2015} with two downsampling and upsampling blocks with 3D convolution layers and skip connections. Hyperparameters are the following: \(t=500\) diffusion steps; linear noise schedule with variances in the range of \(\beta_1 = 10^4\) and \(\beta_t = 0.02\); batch size of 8 and learning rate of 1e-4. The model is implemented using PyTorch~\citep{pytorch_2019} and trained for 200 epochs on 1 GPU NVIDIA Tesla V100. 

The auxiliary classifier used to extract class conditional features contains five 3-dimensional convolution layers with 3-dimensional batch normalization and leaky rectified linear units (ReLU) activation functions, followed by a fully connected layer. The latent features correspond to \(4,096\) flattened vectors injected as conditioning to the U-Net. It is trained for 150 epochs using a learning rate of 1e-4 and a batch size of 64 on 1 GPU NVIDIA Tesla V100.

\subsection*{Evaluation metrics}
We used two types of metrics: Pearson's correlation coefficient and Mean Squared Error (MSE) to study the adequacy of generated images to the ground truth target, and Inception Score~\citep{salimans_improved_2016} (IS) to explore the quality and diversity of the generated images. IS combines the confidence of the class predictions (\textit{i.e.} each image’s label distribution \(p(Y|X)\)) with the variety in the output of the model (\textit{i.e.} the marginal label distribution for the whole set of images \(P(Y)\)). As an additional evaluation criterion, we used the auxiliary classifier to classify the generated images and verify if these images were correctly classified in the target pipeline class. 

\section*{Results} \label{styletransfer:results}

\subsection*{GAN frameworks}

\begin{table}[htb!]
    \centering
    \footnotesize
    \begin{tabular}{p{1.7cm}|c|c|c|c|c}
        \hline
         & & \multicolumn{1}{c|}{fsl-1 $\rightarrow$ spm-0} & \multicolumn{1}{c|}{spm-0 $\rightarrow$ fsl-1} & \multicolumn{1}{c|}{fsl-1 $\rightarrow$ spm-1} & \multicolumn{1}{c}{fsl-1 $\rightarrow$ fsl-0} \\ 
         \hline
         & IS & \multicolumn{4}{|c}{Mean correlations (Std. errors)} \\
         \hline
         \textit{Initial} & \textit{3.69} & \textit{78.2 (0.5)} & \textit{78.2 (0.5)} & \textit{82.8 (0.3)} & \textit{92.3 (0.5)} \\ 
         Pix2Pix & - & \textbf{91.4 (0.1)} & \textbf{89.1 (0.2)} & \textbf{90.1 (0.2)} & \textbf{97.4 (0.1)} \\ 
        CycleGAN & - & 85.5 (0.3) & 67.1 (0.4) & 70.0 (0.5) & 71.2 (0.4) \\ 
        StarGAN & 3.63 & 90.5 (0.4) & 86.8 (0.5) & 87.6 (0.5) & 91.5 (0.3) \\
         \hline
    \end{tabular} \\ 
    \vspace{0.5cm}
    \begin{tabular}{p{1.7cm}|c|c|c|c}
        \hline
         & \multicolumn{1}{c|}{fsl-1 $\rightarrow$ spm-0} & \multicolumn{1}{c|}{spm-0 $\rightarrow$ fsl-1} & \multicolumn{1}{c|}{fsl-1 $\rightarrow$ spm-1} & \multicolumn{1}{c}{fsl-1 $\rightarrow$ fsl-0} \\ 
         \hline
         & \multicolumn{4}{|c}{Mean MSE (Std. errors)} \\
         \hline
         \textit{Initial} & \textit{0.0076 (0.0003)} & \textit{0.0076 (0.0003)} & \textit{0.0041 (0.0002)} & \textit{0.0022 (0.0001)} \\ 
         Pix2Pix & \textbf{0.0027 (0.0001)} & \textbf{0.0014 (0.0001)} & \textbf{0.0025 (0.0001)} & \textbf{0.0005 (0.0)} \\ 
        CycleGAN & 0.0049 (0.0002) & 0.0049 (0.0002) & 0.0072 (0.0002) & 0.0048 (0.0001) \\ 
        StarGAN & 0.0035 (0.0002) & 0.002 (0.0001) & 0.0035 (0.0001) & 0.0017 (0.0001) \\
         \hline
    \end{tabular}
    \caption{Performance of {GAN} frameworks. IS means ``Inception Score". Pearson's correlation (\%) and Mean Squared Error (MSE) are computed between generated and ground truth images and averaged across 20 images. \textit{Initial} represents the metrics between the source image (before transfer) and the ground-truth target image. \textbf{Boldface marks the top model}.}
    \label{tab:styletransfer:results:gan}
\end{table}

In Table~\ref{tab:styletransfer:results:gan}, we show the performance of GAN frameworks for transfers between pipelines with: a different software and a different HRF (columns 1-4), a different software and the same HRF (columns 4-6) and, the same software and a different {HRF} (columns 6-8). Overall, using Pix2Pix~\citep{isola_image--image_2018} and StarGAN~\citep{choi_stargan_2018}, the conversion of statistic maps between pipelines is successful, with increased correlations between target and generated maps compared to correlations between source and target (similar observations are made with MSE, in Supplementary Tables), \textit{e.g.} 91.4\% for target-generated compared to 76.2\% for source-target with Pix2Pix for conversion ``fsl-1'' to ``spm-0''. 

We can point out the large superiority of the supervised Pix2Pix framework compared to unsupervised alternatives. By benefiting from paired data, Pix2Pix generates images closer to the target image than the source image for all transfers. Correlations between target and generated images are close to \(0.9\), which is nearly perfect. On the other hand, the CycleGAN~\citep{zhu_unpaired_2020} framework gives surprising results, relatively low compared to other frameworks. While it makes use of a cyclic loss, similarly to StarGAN~\citep{choi_stargan_2018}, it only learns transfers between two domains. We can suppose that StarGAN leverages the data from the multiple source domains and benefits from the additional classification loss, leading to higher performance in similar settings.

\begin{figure}[hbp!]
    \centering
\includegraphics[width=\textwidth]{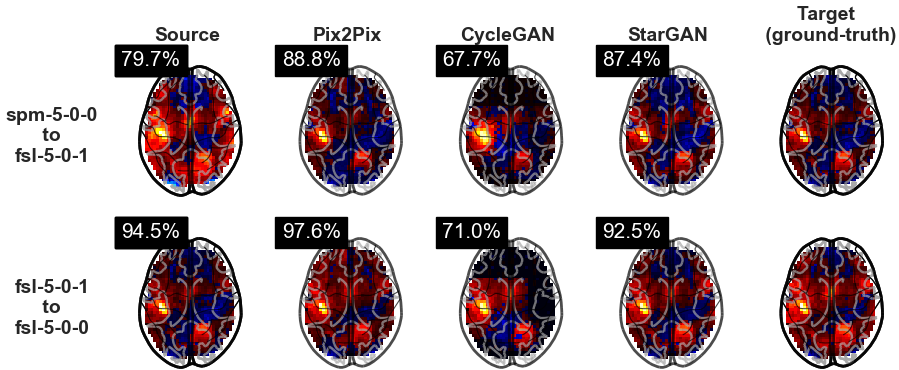}
\caption{Generated images for two transfers and different competitors: Pix2Pix~\citep{isola_image--image_2018}, CycleGAN~\citep{zhu_unpaired_2020} and starGAN~\citep{choi_stargan_2018}. Correlation with ground truth are indicated above each image (in percent). }
    \label{fig:styletransfer:results:gan}
\end{figure}

In Figure~\ref{fig:styletransfer:results:gan}, we illustrate two transfers: (first row) between pipelines with different software packages and different HRF (spm-0 to fsl-1) and (second row) between pipelines with the same software package and different HRF (fsl-1 to fsl-0). We randomly selected a statistic map of the source pipeline and generated the corresponding converted map in the target pipeline. We also display the ground-truth map in the target pipeline. Maps generated using Pix2Pix are closer to the ground truth, with more similar patterns, as seen with the similarity metrics. 

\subsection*{{DM} frameworks}

\begin{table}[htb!]
    \centering
    \footnotesize
    \begin{tabular}{p{1.45cm}|c|c|c|c|c} 
        \hline
         & & \multicolumn{1}{c|}{fsl-1 $\rightarrow$ spm-0} & \multicolumn{1}{c|}{spm-0 $\rightarrow$ fsl-1} & \multicolumn{1}{c|}{fsl-1 $\rightarrow$ spm-1} & \multicolumn{1}{c}{fsl-1 $\rightarrow$ fsl-0} \\ 
         \hline
         & IS & \multicolumn{4}{|c}{Mean correlations (Std. errors)} \\
         \hline
         \textit{Initial} & \textit{3.69} & \textit{78.2 (0.5)} & \textit{78.2 (0.5)} & \textit{82.8 (0.3)} & \textit{92.3 (0.5)} \\ 
         One-hot & 3.66 & 83.9 (0.7) & 75.0 (0.9) & 78.8 (0.8) & 81.1 (0.6) \\ 
         N=1 & 3.70 & 85.4 (0.6) & 77.4 (0.8) & 80.1 (0.8) & 82.8 (0.8) \\ 
         N=10, \(\infty\) & \textbf{3.86} & \textbf{86.1 (0.4)} & \textbf{78.9 (0.6)} & \textbf{81.5 (0.4)} & \textbf{84.1 (0.6)}  \\
         \hline         
    \end{tabular} \\
    \vspace{0.5cm}
    \begin{tabular}{p{1.45cm}|c|c|c|c}
        \hline
         & \multicolumn{1}{c|}{fsl-1 $\rightarrow$ spm-0} & \multicolumn{1}{c|}{spm-0 $\rightarrow$ fsl-1} & \multicolumn{1}{c|}{fsl-1 $\rightarrow$ spm-1} & \multicolumn{1}{c}{fsl-1 $\rightarrow$ fsl-0} \\ 
         \hline
         & \multicolumn{4}{|c}{Mean MSE (Std. errors)} \\
         \hline
         \textit{Initial} & \textit{0.0076 (0.0003)} & \textit{0.0076 (0.0003)} & \textit{0.0041 (0.0002)} & \textit{0.0022 (0.0001)} \\ 
         One-hot & 0.0097 (0.0014) & 0.0048 (0.0007) & 0.0088 (0.0014) & 0.0043 (0.0003) \\ 
         N=1 & 0.0053 (0.0003) & 0.0037 (0.0003) & 0.0073 (0.0009) & 0.0037 (0.0003) \\ 
         N=10, \(\infty\) & \textbf{0.0043 (0.0003)} & \textbf{0.0028 (0.0002)} & \textbf{0.0049 (0.0003)} & \textbf{0.0029 (0.0002)}  \\
         \hline         
    \end{tabular}
    \caption{Performance of DM frameworks. IS means ``Inception Score". Pearson's correlation (\%) and Mean Squared Error (MSE) are computed between generated and ground truth images and averaged across 20 images per transfer. \textit{Initial} represents the metrics between the source image (before transfer) and the target image. \textbf{Boldface marks the top model}.}
    \label{tab:styletransfer:results:ddpm}
\end{table}

In Table~\ref{tab:styletransfer:results:ddpm}, we show the performance of the DM frameworks for the same four transfers. Different frameworks are compared: one-hot conditioning from~\cite{ho_classifier-free_2022}, auxiliary classifier-conditioning with \(N=1\) target image selected randomly, inspired from~\cite{preechakul_diffusion_2022}, and auxiliary classifier-conditioning with \(N=10\) target images selected randomly (named \(N=10, \infty\) in the Table). 

The conversion between pipelines seems more difficult than with the GAN frameworks. While all models succeed in changing the class identified by a pipeline classifier to the target domain, the success of the conversion in terms of similarity to the ground truth is variable across transfers. For instance, all DM frameworks succeed for the transfer ``fsl-1'' to ``spm-0'', while none is successful for the transfer ``fsl-1'' to ``fsl-0''. These low performance could be explained by the difficulty of the models to learn differences between pipelines that provide similar results (\textit{i.e.} whose results display very similar activation patterns).  

Using a DM with auxiliary classifier conditioning and multiple target images (\(N=10, \infty\)) improves the performance compared to the alternative frameworks. Both the quality and diversity of images are increased (\(IS=3.86\)). In terms of similarity to the ground truth, this framework also outperforms other DM models by up to 4\% in correlations between ground-truth and generated image compared to~\cite{ho_classifier-free_2022} for the transfer ``spm-0'' to ``fsl-1'' and up to 3\% for ``fsl-1'' to ``spm-0''. 

\begin{figure}[htbp!]
    \centering
    \includegraphics[width=\textwidth]{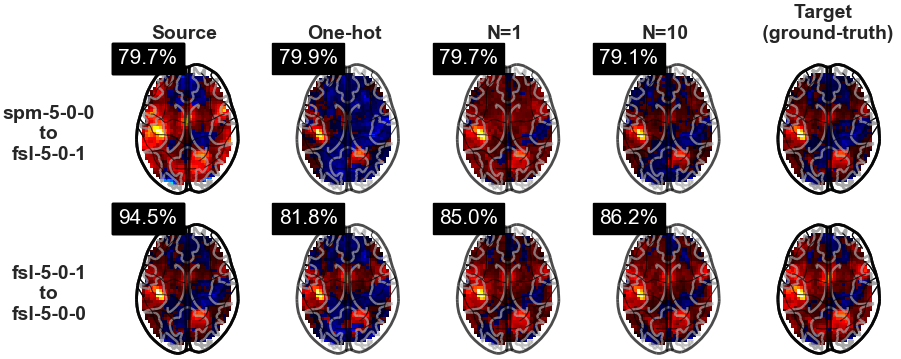}
    \caption{Generated images for two transfer and different competitors: conditioning with one-hot encoding~\citep{ho_classifier-free_2022}, with a classifier-conditioning N=1 and N=10 target images randomly selected. Correlations with ground truth are indicated above generated and source images (in percent). }
    \label{fig:styletransfer:results:ddpm}
\end{figure}

The first row of Figure~\ref{fig:styletransfer:results:ddpm} illustrates a transfer between pipelines with different software packages and different HRF (``spm-0'' to ``fsl-1''). The second row shows a transfer between pipelines with the same software package and different HRF (``fsl-1'' to ``fsl-0''). DM with multiple target images generates statistic maps close to the ground truth for both transfers, representing the intrinsic properties of the map while modifying its extrinsic properties to the target domain. Using the one-hot encoding, generated statistic maps seem far from the ground truth, failing to represent the full characteristics of the target domain.

The performance of DM frameworks remains notably inferior to the ones obtained with Pix2Pix~\citep{isola_image--image_2018} or StarGAN~\citep{choi_stargan_2018}. This superiority can be explained by the differences between frameworks: GAN methods use adversarial training and StarGAN improves this by using a classifier loss and a cyclic-reconstruction loss. Moreover, the sampling process of GAN relies on the source image directly and does not require setting an initial state, which might facilitate the source content preservation. However, we can note that Inception Scores (IS) obtained with DM frameworks are better than the ones obtained with StarGAN, which indicates that images generated by DM frameworks are more diverse. This observation is consistent with the literature \citep{dhariwal_diffusion_2021} and the sampling process of DM frameworks which includes randomness. 

\section*{Conclusion}\label{styletransfer:conclu}

In this study, we explored the potential of style transfer frameworks on the task of converting fMRI statistic maps between different pipelines. We showed that the StarGAN framework, trained on unsupervised and multi-domain data, could be easily trained and applied to generate statistic maps that maintain the intrinsic properties of brain activity while changing the style of the image. These newly generated data could be used to build valid mega-analyses on heterogeneous datasets and hence increase sample sizes in fMRI data analysis. Future works will focus on validating these assumptions by evaluating the false positives and false negatives rates of mega-analyses built with converted statistic maps. 

\clearpage  
\midlacknowledgments{
This work was supported by {Region Bretagne (ARED MAPIS)} and {ANR project ANR-20-THIA-0018)}. 
Data were provided by the Human Connectome Project, WU-Minn Consortium (Principal Investigators: David Van Essen and Kamil Ugurbil; 1U54MH091657) funded by the 16 NIH Institutes and Centers that support the NIH Blueprint for Neuroscience Research; and by the McDonnell Center for Systems Neuroscience at Washington University.}

\section*{Data statement}\label{styletransfer:datastatement}

This study was performed using derived data from the HCP Young Adult~\citep{van_essen_wu-minn_2013}, publicly available at ConnectomeDB. Data usage requires registration and agreement to the HCP Young Adult Open Access Data Use Terms available at:~\cite{hcp_dua}. 

The HCP multi-pipeline dataset~\citep{germani_hcp-multi-pipeline_2024} is publicly  available on Public-nEUro~\citep{public-neuro}: \url{https://publicneuro-catalogue.netlify.app/dataset/PN000003%20HCP%20multipipelines/V1}. 

\section*{Ethics}
This study was performed using derived data from the HCP Young Adult~\citep{van_essen_wu-minn_2013}.  No experimental activity involving the human participants was made by the authors. Only shared data were used.

Written informed consent was obtained from participants and the original study was approved by the Washington University Institutional Review Board. 

We agreed to the HCP Young Adult Open Access Data Use Terms available at:~\cite{hcp_dua}.

\section*{Code and data availability}\label{styletransfer:code}

All the scripts used to perform the study (models training, testing and performance evaluation) are available on Software Heritage: \href{https://archive.softwareheritage.org/swh:1:snp:b0b52aa88bef8f4411bdd7e00a2d71715d7830bb;origin=https://github.com/elodiegermani/style-transfer_diffusion}{swh:1:snp:b0b52aa88bef8f4411bdd7e00a2d71715d7830bb} \citep{code_SH}.

Derived data such as pre-trained models and computed metrics are available on Zenodo~\citep{germani_2024_13748563}.

\bibliography{bibliography}

\newpage
\appendix
\section*{Supplementary materials}
\subsection*{Generalizability study}

Ideally, researchers should be able to re-use a model trained to convert statistic maps of a task for another one. Thus, we test the generalizability of the frameworks trained to convert statistic maps from a particular task to statistic maps of other tasks. Here, we explore the generalizability of StarGAN trained on maps of \textit{right-hand} to maps of \textit{right-foot} and \textit{left-hand}. 

\begin{table}[htbp]
\footnotesize
    \centering
    \begin{tabular}{p{3.4cm}|c|c|c|c}
    & fsl-1 $\rightarrow$ spm-0 & spm-0 $\rightarrow$ fsl-1 & fsl-1 $\rightarrow$ spm-1 & fsl-1 $\rightarrow$ fsl-0 \\ 
    \hline
    \multicolumn{5}{c}{\textbf{Converting \textit{right-foot} statistic maps with:}} \\
    \hline
    \textit{Initial} & \textit{86.2} & \textit{86.3} & \textit{85.8} & \textit{95.9} \\
    Trained on \textit{right-foot} & 90.5 & 88.8 & 88.9 & 93.1 \\
    Trained on \textit{right-hand} & 71.2 & 71.0 & 63.0 & 82.2 \\
    \hline 
    \multicolumn{5}{c}{\textbf{Converting \textit{left-hand} statistic maps with:}} \\
    \hline
    \textit{Initial} & \textit{82.3} & \textit{82.3} & \textit{85.9} & \textit{92.9} \\
    Trained on \textit{left-hand} & 88.4 & 85.4 & 86.8 & 85.8 \\ 
    Trained on \textit{right-hand} & 73.2 & 74.5 & 69.5 & 75.2 \\
    \end{tabular} \\ 
    {Supplementary Table 1 - Robustness to distribution shifts (\textit{i.e.} trained and evaluated with statistic maps from different tasks) of StarGAN. Pearson’s correlation (\%) is computed between generated and ground truth and averaged across 20 images per transfer. Initial represents the metrics between the source image (before transfer) and the target image.}\label{tab:styletransfer:results:generalizability}
\end{table}

Supplementary Table 1 shows the performance of the StarGAN framework trained on \textit{right-hand} statistic maps when applying it to statistic maps of \textit{right-foot} and \textit{left-hand}. We also show the performance of StarGAN trained directly on \textit{right-foot} and \textit{left-hand} for comparison. Our results show that the StarGAN framework does not behave positively when applied to a task different from the task used in the training data: the generated images seem further from the target images (\textit{e.g.} 71.2\% for the first transfer between ``fsl-1'' to ``spm-0'') than the source image was from the target (\textit{e.g.} 86.2\% in this case). We can see a large performance drop between the framework trained on statistic maps from the task compared to the one trained on maps from another task. Similar observations can be made for generalizability to closer tasks (here, frameworks trained on \textit{right-hand} evaluated on \textit{left-hand}). These results make us suppose that the mapping from one pipeline to another is different between tasks and thus, if the same model can be used for different subjects on the same task, different models should be trained to transfer statistic maps from tasks unseen during training.

\newpage
\subsection*{Supplementary figures}

\begin{table}[h!]
    \centering
    \begin{tabular}{c|c|c|c|c}
    \hline
    Pipelines & Layer 1 & Layer 2 & Layer 3 & Layer 4 \\ 
    \hline
    \multicolumn{5}{c}{Same software, different parameters} \\
    \hline
    fsl-5-0-0 / fsl-5-0-1 & 86.5 & 91.4 & 95.4 & 99.2 \\
    spm-5-0-0 / spm-5-0-1 & 86.5 & 90.9 & 94.2 & 98.4 \\
    \hline
    \multicolumn{5}{c}{Same parameters, different software} \\
    \hline
    fsl-5-0-0 spm-5-0-0 & 88.8 & 88.2 & 93.6 & 98.2 \\
    fsl-5-0-1 spm-5-0-1 & 84.8 & 85.8 & 92.4 & 98.0 \\
    \hline
    \multicolumn{5}{c}{Different software, different parameters} \\ 
    \hline
    fsl-5-0-0 spm-5-0-1 & 74.5 & 81.0 & 88.7 & 97.1 \\
    fsl-5-0-1 spm-5-0-0 & 74.8 & 77.7 & 88.2 & 97.3 \\
    \hline
    \end{tabular}
    \\ \vspace{0.2cm}
    \textbf{Supplementary Table S1.} Mean correlations between feature maps learned at each layer for each pair of pipelines. Features are close for pipelines sharing the same software at Layer 4, which might explain the difficulty to rely on these features to perform transfer. 
\end{table}

\begin{figure}[h!]
    \centering
    \includegraphics[width=0.8\textwidth]{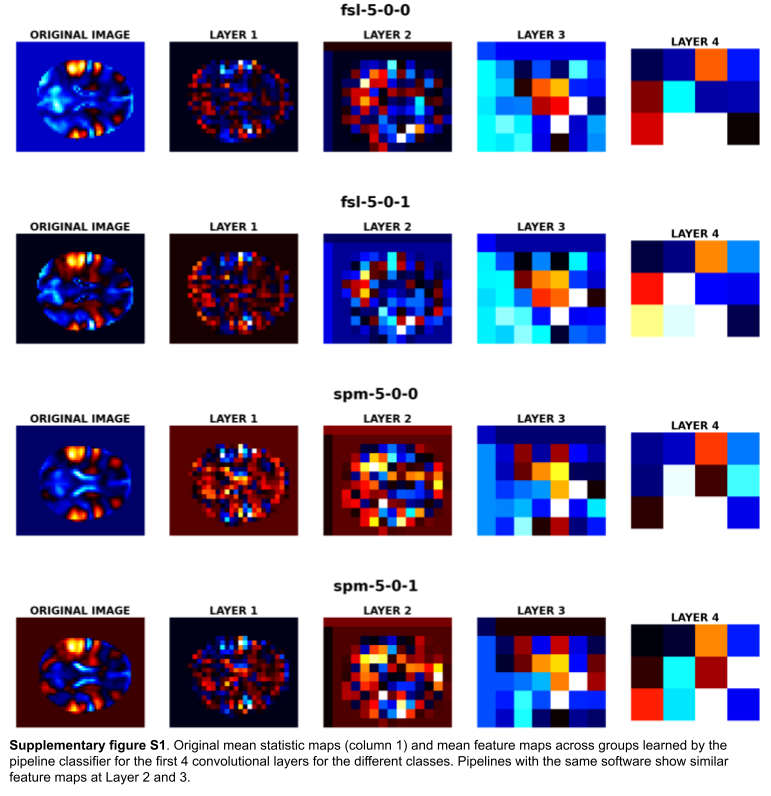} 
\end{figure}

\begin{figure}[htbp!]
    \centering
    \includegraphics[width=\textwidth]{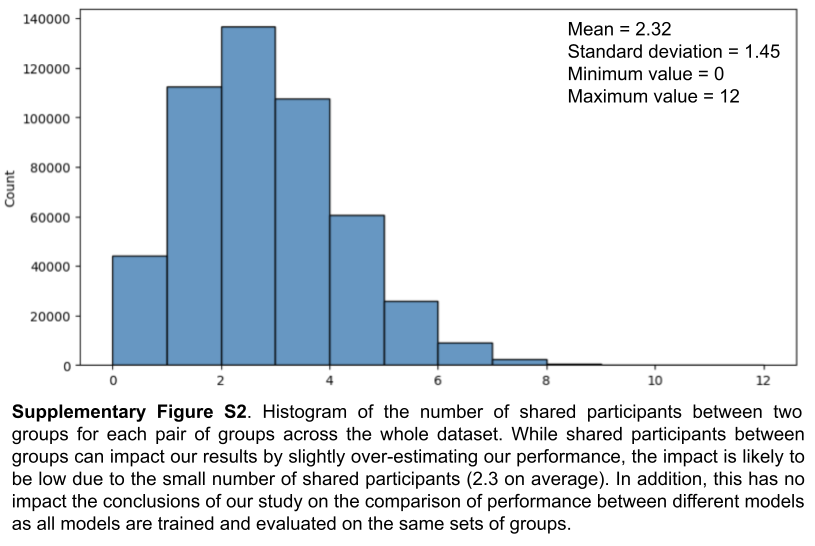}
\end{figure}

\end{document}